\documentstyle[epsfig,aps,pre]{revtex}
\tighten
 
\begin{document}
\draft
\twocolumn
 
\wideabs{
\title{Unfolding Rates for the Diffusion-Collision Model}
 
\author{Chris Beck and Xavier Siemens}
 
\address{Molecular Modeling Laboratory,
Department of Physics and Astronomy,
Tufts University,
Medford MA 02155, USA}
 
\date{\today}
 
\maketitle
 
\begin{abstract}
In the diffusion-collision model, the unfolding rates are given
by the likelihood of secondary structural cluster dissociation. In this work,
we introduce an unfolding rate calculation for proteins whose secondary 
structural elements are $\alpha$-helices, modeled from thermal
escape over a barrier which arises from the free energy in 
buried hydrophobic residues. Our results are in good agreement with currently
accepted values for the attempt rate.
\end{abstract}

 
}

\section{Introduction}

In the diffusion-collision model of protein folding \cite{1} the protein is modeled 
using a collection of spheres connected by flexible strings. The spheres 
represent the secondary
structural elements such as \( \alpha  \)-helices or \( \beta  \)-sheets 
(or clusters of these secondary structures)
called microdomains, that constitute the protein. 

The folding process from a completely unfolded protein to the the final native
state is accomplished via diffusion through the solvent, collision, and finally
coalescence of the microdomains. The state of the protein is defined by the
number of pairings between the microdomains that
are present at a given time \( t \). The rate equations for transitions between
these states can be written as
\begin{equation}
\label{rateeqn}
\frac{d{\bf P}(t)}{dt}={\hat{K}}{\bf P}(t)
\end{equation}
where \( {\bf P}(t) \) is the vector of states and \( {\hat{K}} \) is a matrix
containing the transition rates between the different states. A protein having, 
say, \( q \) microdomains would involve
\( p=q(q-1)/2 \) pairings, \( 2^{p} \) states \( P_{i}(t) \) and a \( 2^{p}\times 2^{p} \)
rate matrix \( \hat{K} \). 

In general, the calculation of the elements of the
rate matrix \( {\hat{K}} \) is somewhat involved. The forward rates are the
rates of structural coalescence. In the diffusion-collision model the forward
rates are calculated assuming the microdomains 
diffuse through a solvent environment,
the space of which is limited by the length of the intervening strings and the
van der Waals radii of the microdomains. These microdomains are assumed to be
nascently formed, and their degree of formation is given by a helix-coil transition
theory calculation \cite{2} (as in AGADIR \cite{3,4}) in the case of \( \alpha  \)-helices,
or via a combination of theory \cite{5} and experiment \cite{6}
in the case of \( \beta  \)-sheets. As the microdomains undergo diffusion,
they occasionally collide. When this happens the microdomains coalesce with
a probability \( \gamma  \), being held together by hydrophobic interactions
in the case of \( \alpha  \)-helices, or a combination of hydrophobic and hydrogen
bond interactions in the case of \( \beta  \)-sheets. The coalescence probability
\( \gamma  \) is given by the likelihood that the microdomain is in \( \alpha  \)-helical
or \( \beta  \)-sheet form, the percentage of hydrophobic area, and the likelihood
of proper geometrical orientation upon collision.

The forward folding times between any two given states
in the mean first passage time approximation \cite{7} are given by
\begin{equation}
\label{MFPT}
\tau _{f}=\frac{l^{2}}{D}+\frac{LV(1-\gamma )}{\gamma DA}
\end{equation}
where \emph{\( V \)} is the diffusion volume available to the microdomain
pair, \emph{\( A \)} is the target area for collisions, \emph{\( D \)} is
the relative diffusion coefficient, \( \gamma  \) is the probability of coalescence
upon collision and \emph{\( l \)} and \emph{\( L \)} are geometrical parameters
calculated for diffusion in a spherical space. The inverse of the first passage
time-scales \( \tau _{f} \) are the forward folding rates $ k_{f} $
that are used in the rate matrix \( {\hat{K}} \).

The pairs can also dissociate. In typical diffusion-collision
model calculations, the form of the unfolding times \( \tau _{b} \)
used for two microdomains A and B comes from 
the Van't Hoff-Arrhenius law given by
\begin{equation}
\label{brDCM}
\tau _{b}=\nu ^{-1}e^{\Delta G_{AB}/k_{B}T}
\end{equation}
where $\Delta G_{AB}$ is the free energy difference between paired and unpaired states,
\emph{\( k_{B} \)} is Boltzmann's constant, \emph{\( T \)} is the temperature 
and \( \nu  \) is an attempt rate. In the case of $\alpha$-helices the dominant 
contribution to the free energy comes from the buried hydrophobic area and 
therefore
\begin{equation}
\label{alphahel}
\Delta G_{AB}=f A_{AB}
\end{equation}
where \emph{\( f \)} is the free energy change per unit buried hydrophobic
area in the pairing \cite{9} and \emph{\( A_{AB} \)} is the buried area \cite{10}.
The unfolding rates $k_b$ are given by the inverse of the unfolding times $\tau_b$.

The diffusion-collision model has been successful in describing 
the overall folding kinetics of several proteins  \cite{11,Meyers,rohit}. 
In each of these studies a single value of the 
parameter $\nu$ was used for every unfolding transition. 
This value was adjusted to obtain the desired result, namely, 
to ensure that the protein would fold to its native state. This 
procedure is justified because the equilibrium (or native) 
occupation probabilities are 
known; in fact, for sufficiently simple systems 
the folding and unfolding rates can be 
determined from 
these probabilities \cite{chandler}. The typical values used lie 
between $1$-$1000\ ns^{-1}$, which yields unfolding rates consistent 
with observed rates of bi-molecular dissociation \cite{12}.
   
In cases where the final occupation probabilities are unknown, for
instance in studies of protein mis-folding and non-native kinetic
intermediates \cite{14} such methods are clearly not possible.
Indeed, even a detailed description of the intermediate folding kinetics 
of a protein whose final state {\it is} known requires a more
accurate and foundational determination of $\nu$, as was pointed out by
Burton et al. \cite{11}.
   
In this work we compute unfolding rates that, in the context of the 
diffusion-collision model, can be used for any given unfolding 
transition in the study of
proteins whose secondary 
structural elements are $\alpha$-helices. From the rates we find the values of the parameter
$\nu$. This makes the 
diffusion-collision model more predictive and enables it to be used in 
situations where the occupation probabilities are unknown.

\section{Calculation of the Unfolding Rates}

\begin{figure}
\begin{center}
\includegraphics[width=5cm]{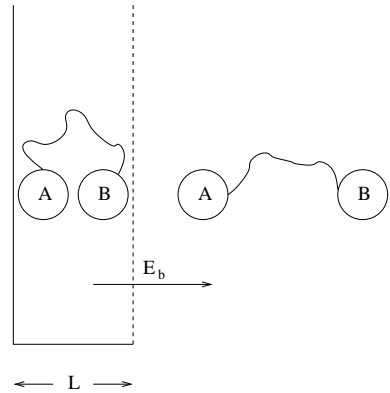}
\end{center}
\caption{Potential for the two microdomains A and B. 
The potential is infinite on the left because
of the hard-core repulsion of the van der Walls contact between the microdomains.
The barrier on the right can be crossed by microdomain pairs 
with energies larger
than
\protect\( E_{b} = fA_{AB}\protect \),
the free energy difference between paired and unpaired states with a buried
hydrophobic area \protect\( A_{AB}\protect \). The width of the well \protect\( L\protect \),
is taken to be the diameter of a water molecule.} 
\label{fig:abc}
\end{figure}

We model the dissociation of microdomains as a thermal escape event over
a barrier. Consider the pair of microdomains 
(which could be $\alpha$-helices or clusters of $\alpha$-helices)  
A and B connected by a string, diffusing in the potential well 
depicted in Figure \ref{fig:abc}. The left boundary is infinite because of the 
hard-core repulsion of the
van der Waals contact between the pair. 
Pairs with energies  larger than \( E_{b} = fA_{AB} \), 
the free energy difference between paired and unpaired states, can escape 
from the right boundary of the well.
The
well width \( L \) is set to the diameter of a water molecule. A separation
larger than \( L \) exposes the buried hydrophobic area of 
the pair to the solvent, the free energy
savings is lost, and the pair separates, resulting in an escape from
the potential well.

The binding energies $E_b$ of microdomain pairs in proteins
are typically much larger than the thermal energy
\begin{equation}
\label{EbggkT}
E_b \gg k_B T.
\end{equation}
This means that the time to escape from the well is much larger than any other 
time-scale involved in the problem, in particular larger than the thermalisation 
(or velocity auto-correlation) time and larger than the 
time it takes for the pair to diffuse in the well. Consequently, at any one time, 
the spatial distribution inside the well of an ensemble of pairs will 
be homogeneous 
\begin{equation}
\label{rho}
\rho(x,t) \propto 1/L
\end{equation}
and the flux incident on the barriers will be thermal. We will use these 
two facts to calculate the rate at which the pairs dissociate.

The flux at the boundary on the right (at \( x=L \)) depends on the 
density of pairs
at that boundary and the probability that their energy is high enough
to thermally escape over the boundary. The differential element of flux at
the boundary \( L \) of pairs with relative velocity between \emph{\( v \)}
and \emph{\( v+dv \)} is given by
\begin{equation}
\label{bc@L}
dJ^{out}(L,t)=v\rho (L,t)dN(v)
\end{equation}
where $\rho (L,t)$ is the number density of pairs at the boundary at a time $t$,
\begin{equation}
\label{MBDist}
dN(v)=\left( \frac{\mu }{2\pi k_{B}T}\right) ^{1/2}e^{-\mu v^{2}/2k_{B}T}dv
\end{equation}
is the fraction of pairs with
relative velocities between \( v \) and \( v+dv \), and \( \mu  \) is the
reduced mass given by
\begin{equation}
\label{reducedM}
\mu =\frac{m_{A}m_{B}}{m_{A}+m_{B}}
\end{equation}
where \( m_{A} \) and \( m_{B} \) are the masses of the two microdomains.

In order to find the total flux through the outer boundary
at \emph{\( L \)} we must integrate over all velocities larger
than \( +\sqrt{E_{b}/2m} \) since the potential barrier can be crossed 
by pairs with energies higher than \( E_{b} \),
and pairs with relative velocities higher than that can contribute
to the flux leaving the well. This yields a flux out of the well
\begin{equation}
\label{totalJ@L}
J^{out}(L,t)=\rho (L,t)\left( \frac{k_{b}T}{2\pi \mu }\right) ^{1/2}e^{-E_{b}/k_{B}T}
\end{equation}
If the number of pairs inside the well at some time $t$ is $n(t)$ then, 
because of (\ref{EbggkT}), the number density must be
\( \rho (x,t)=n(t)/L \) everywhere and 
\begin{equation}
\label{totalJ@L2}
J^{out}(L,t)=\frac{n(t)}{L}\left( \frac{k_{b}T}{2\pi \mu }\right) ^{1/2}e^{-E_{b}/k_{B}T}.
\end{equation}
This means that the dissociation rate constant for a pair of microdomains
with reduced mass $\mu$ and buried hydrophobic
area $A_{AB}=E_b/f$ at a temperature $T$ is
\begin{equation}
\label{fluxovereprob}
k_b=\frac{1}{L}\left( \frac{k_{B}T}{2\pi \mu }\right) ^{1/2}e^{-E_{b}/k_{B}T}.
\end{equation}
The terms preceding the exponential correspond to our prediction for the Van't
Hoff-Arrhenius attempt rate \( \nu  \) in (\ref{brDCM}). As an example, the
attempt rate found for a coalesced pair of 16-residue Regan-Degrado \cite{15}
helices with a combined hydrophobic area loss of \( 600 \)\AA\( ^{2} \) is
\( 64\times 10^{9}s^{-1} \). 

It is interesting to note that a result similar to (\ref{fluxovereprob}) would have been
obtained by assuming the the attempt rate to be the inverse of the thermal well-crossing 
time, namely taking
\begin{equation}
\label{nutherm}
\nu \sim {1 \over L} \sqrt{{k_B T \over \mu}}
\end{equation}
in (\ref{brDCM}). This is, in fact, not the origin of the pre-factors 
in (\ref{fluxovereprob}). 
They arise as 
a consequence of (\ref{EbggkT}): The factor of  $1/L$ comes from 
the homogeneity of the spatial distribution (\ref{rho}) and the factor 
of $\sqrt{{k_{B}T}/{2\pi \mu }}$ from the integration of the thermal 
velocity distribution (\ref{MBDist}).

It is possible that dissociation events within a protein also include 
a relative rolling and/or sliding motion of the microdomains. In this case the 
calculation above can be performed with a few minor
differences that take into account the extra degrees of freedom. 
The relative velocity distribution of the microdomains is still
the one-dimensional Maxwell-Boltzmann distribution because motion
parallel to the surface through which the probability is flowing 
does not contribute to escape from the well. The probability in the bound region is
homogeneously distributed in a two- or three-dimensional volume
in (\ref{totalJ@L}), and flows out of that volume through
a one- or two-dimensional area. This calculation yields the result
\begin{equation}
\label{2Dbackrate}
k_b=\frac{d}{L}\left( \frac{k_{b}T}{2\pi \mu }\right) ^{1/2}e^{-E_{b}/k_{B}T},
\end{equation}
where we set $d=2$ if we include either the rolling {\it or} sliding degrees 
of freedom, and $d=3$ if both of them are included. Due to the steric clashing of the 
side chains it seems rather unlikely that dissociation would include a 
sliding motion along the axes of the microdomains.
It may be relevant, however, in the context of molten globules.

This approach succeeds in removing the free parameter \( \nu  \) from the diffusion-collision
model. Moreover, our results for the one-, two- and three-dimensional unfolding rates 
have a \( \sqrt{T/\mu } \) dependence that could be used to distinguish 
between this and other proposals for the mechanism of microdomain 
pair dissociation.

\section{Concluding Remarks}

We have presented a calculation for the dissociation rate of a microdomain 
pair using
a simple potential barrier over which pairs having energies above the free energy 
of the hydrophobic docking can escape. Since we have not accounted for the energy in hydrogen 
bonds this result is relevant for the dissociation of $\alpha$-helix pairs or helix 
cluster pairs only and not for the dissociation of $\beta$-sheet pairs. 
We have found the unfolding rates arising from thermal 
fluctuations out of this potential well to be in good agreement with currently accepted 
values of the attempt rate \( \nu  \). 

The motivation of this work was to eliminate the free parameter \( \nu  \)
from the diffusion-collision model. In previous applications of
the diffusion-collision model (see for example \cite{11,Meyers,rohit}) 
the folding kinetics from a denatured or random
coil state to the final native state were followed. In such a case, it is reasonable
to set the parameter \( \nu  \) such that the native state achieves most of
the probability, because we know that the final state is attained
at the end of the folding process. However, the removal of this parameter is important 
when considering folding processes which do not involve the native state. 
For example, in studying intermediate processes or 
protein mis-folding \cite{14}, where the occupation
probabilities may be completely unknown, such reasonable estimates of \( \nu  \)
are not available. In these cases, elimination of \( \nu  \) as a free parameter
is crucial.

The results presented here also predict a \( \nu \propto \sqrt{T/\mu } \) dependence
in all cases which can be distinguished experimentally from other possibilities
such as  \( \nu \propto T  \) \cite{12}.
Another difference is the dependence of the unfolding rates on the states,
not only through the hydrophobic area, but also through the reduced mass \( \mu  \)
of the microdomains undergoing dissociation. This
is markedly different from typical diffusion-collision model calculations where
the attempt rate \( \nu  \) is assumed to be the same for all dissociation
events within the protein.

\begin{acknowledgments}
We would like to thank Ken Olum, David Weaver, and Larry Ford for many useful
discussions. 
\end{acknowledgments}

\end{document}